\documentclass[12pt,preprint]{aastex}
\shorttitle{SN~2005hj}
\shortauthors{Quimby et al.}
\usepackage{epsfig}
\begin{document}

\title{SN~2005hj: Evidence for Two Classes of Normal-Bright SNe~Ia and Implications for Cosmology\footnotemark[1]}

\author{
  Robert Quimby\altaffilmark{2},
  Peter H\"{o}flich\altaffilmark{3},
  J. Craig Wheeler\altaffilmark{2}
}
\footnotetext[1]{Based on observations obtained with the Hobby-Eberly
Telescope, which is a joint project of the University of Texas at
Austin, the Pennsylvania State University, Stanford University,
Ludwig-Maximilians-Universit\"{a}t M\"{u}nchen, and
Georg-August-Universit\"{a}t G\"{o}ttingen.}
\altaffiltext{2}{Department of Astronomy, University of Texas, Austin, TX 78712, USA}
\altaffiltext{3}{Department of Physics, Florida State University, Tallahassee, FL 32312, USA}

\begin{abstract}
HET Optical spectra covering the evolution from about 6 days before to
about 5 weeks after maximum light and the ROTSE-IIIb unfiltered light
curve of the ``Branch-normal'' Type Ia Supernova SN~2005hj are
presented.  The host galaxy shows \ion{H}{2} region lines at redshift
of $z=0.0574$, which puts the peak unfiltered absolute magnitude at a
somewhat over-luminous $-19.6$. The spectra show weak and narrow
\ion{Si}{2} lines, and for a period of at least 10 days beginning
around maximum light these profiles do not change in width or depth
and they indicate a constant expansion velocity of $\approx 10,600$ km
s$^{-1}$. Our observations indicate that Si is confined to a relatively
narrow velocity region. We analyzed the observations based on detailed
radiation dynamical models in the literature. The models considered
include delayed detonations, deflagrations, pulsating delayed
detonations, and tamped detonation/merger scenarios. Whereas the first
two classes of models have been used to explain the majority of
SNe~Ia, they do not predict a long velocity plateau in the \ion{Si}{2}
minimum with an unvarying line profile. Pulsating delayed detonations
and merger scenarios form shell-like density structures with
properties mostly related to the mass of the shell, $M_{shell}$, and
we discuss how these models may explain the observed \ion{Si}{2} line
evolution; however, these models are based on spherical calculations
and other possibilities may exist. SN~2005hj is consistent with
respect to the onset, duration, and velocity of the plateau, the peak
luminosity and, within the uncertainties, with the intrinsic colors
for models with $M_{shell}=0.2 M_\odot$. Our analysis suggests a
distinct class of events hidden within the Branch-normal SNe~Ia. If
the predicted relations between observables are confirmed, they may
provide a way to separate these two groups.  We discuss the
implications of two distinct progenitor classes on cosmological
studies employing SNe~Ia, including possible differences in the peak
luminosity to light curve width relation.

\end{abstract}


\section{Introduction}

Type Ia supernovae (SNe~Ia) are thought to be thermonuclear
disruptions of white dwarf (WD) stars \citep{hf60}, but the details
remain uncertain.  One possibility for the progenitors is the single
degenerate model in which main sequence stars or post main sequence
red giants transfer mass to a WD through Roche Lobe overflow or a
common envelope and the WD grows close to the Chandrasekhar mass
$M_{Ch}$. Due to compression, the thermonuclear runaway starts near
the center leading to the explosion of the WD and a rapidly expanding
envelope with a mass close to $M_{Ch}$.  A second possibility is the
double degenerate model in which a pair of WDs merge and lead to an
explosion. In most such cases, the resulting mass of the rapidly
expanding envelope will be different from $M_{Ch}$. Residual material
from these mergers surrounding the explosions will get swept up by the
ejecta forming dense, shell-like structures.  Although the explosion
of single WDs seems to be favored for the majority of objects, we may
expect mergers to contribute to the SNe~Ia population (see \S
\ref{obs}).

The possibility of different progenitor channels, the population of
which may vary with redshift, may pose a challenge for the use of
SNe~Ia in cosmological studies that rely on a single parameterization,
such as a light curve width to peak luminosity relation, LWR, to
reduce the intrinsic scatter in the peak magnitudes and render them
standard candles \citep{phillips1993,perlmutter1997}.  To first order,
the LWR relation can be understood as a result of different amounts of
$^{56}$Ni produced during the explosion
\citep{hoeflich1996,umeda99}. There may be some spread and an offset
in LWR introduced by one of the channels if the masses of the envelope
differ from $M_{Ch}$, and/or the density structures differ.  This can
lead to a systematic shift of LWR with redshift if the evolutionary
time scales of the progenitor systems differ.  Even if the different
progenitor scenarios obey the same LWR, differences in the color could
introduce systematic errors in cosmological studies because SNe~Ia are
known to suffer to some degree from reddening in their respective host
galaxies which has to be taken into account. To correct for this, the
maximum light color excess (usually $E_{B-V}$) and an average reddening
law are used to determine the amount of absorption. SNe~Ia that are
intrinsically redder as compared to the average local sample will thus
be over-corrected in this fashion to a higher luminosity.

Similarly to the two distinct progenitor channels, qualitative
variations in the explosion physics may lead to various classes of
SNe~Ia even within the single degenerate scenarios. Standard explosion
models include delayed detonations (DD) and deflagrations. In these
scenarios, burning during the deflagration phase leads to an unbound
WD. In DD models, the deflagration turns into a detonation in an
expanding envelope. Because the density structure of the WD declines
monotonically with radius, the resulting density structure in the
expanding envelope also smoothly declines with mass and radius.  A
variation of DD models are the pulsating delayed detonation models
(PDD; \citealt{khokhlov1993,hoeflich1995}).  In these models, the
total energy production during the deflagration phase is, by
construction, lower and insufficient to unbind the WD. This results in
large amplitude pulsations.  Because the fall-back time increases with
distance, the inner regions contract and leave a shell behind at
larger distances. Due to infall driven compression, a detonation is
triggered, the material behind the burning front is accelerated, and
this expanding material runs into the low-velocity material left at
larger distances. Similar to the merger scenario, a shell-like
structure is formed with very similar light curve and spectroscopic
properties, but with a total mass close to $M_{Ch}$
\citep{hoflich_khokhlov1996}.

These two groups, consisting of DD and deflagration models such as W7
\citep{nomoto1984}, which lack shells, and the models with shells
(mergers and PDDs), can be differentiated by their predictions for the
photospheric evolution and maximum light colors \citep{khokhlov1993,
hoflich_khokhlov1996}. For the former group, the photospheric
velocities, $v_{ph}$, smoothly decline with time and the models show a
blue color at maximum light, $B-V \approx 0^m$; in the latter group,
$v_{ph}$ shows a plateau in the evolution as the photosphere recedes
through the shell. These models are intrinsically redder and slightly
over-luminous because of the lower expansion rate in the inner region.
As shown in \citet{khokhlov1993} the color, length and velocity of the
plateau are correlated with the mass of the shell, and this
potentially allows the two groups to be distinguished even for similar
brightnesses.

Indeed, there is a growing sample of SNe~Ia showing photospheric
velocity plateaus (e.g.  1990N;
\citealt{leibundgut1991,mueller_hoeflich1994}; 1991T, 1999aa;
\citealt{garavini2004}; 1999ee; \citealt{hamuy2002}; 2000cx;
\citealt{li2001}; see also \citealt{benetti2005} ).  Many of these
SNe~Ia have been reported as having a red color $B-V$ at maximum, but
this is typically attributed to reddening along the line of
sight. Alternatively, this sample may suggest the contribution of
events with shell-like density structures in the observed
population. These events may be understood in terms of mergers or
PDDs; however, the inhomogeneities and incompleteness of individual
data sets in the literature preclude definite conclusions.

To address this problem and others, we started the Texas Supernovae
Search (TSS; Quimby et al. in prep.) with the goal of providing a
homogeneous set of quality data for several supernovae beginning well
before maximum light.  In this paper, we present our observations of
SN~2005hj and analysis of the data.  In \S{\ref{obs}} we describe the
discovery and give the details for both the photometric and
spectroscopic follow-up. In \S \ref{models} we discuss generic
properties of explosion models and suggest a secondary parameter to
separate models with and without shells, and analyze the peculiarities
of SN~2005hj.  Conclusions and discussion are presented in
\S{\ref{conclusions}}.

\section{Observations}\label{obs}

\begin{figure}
\epsfig{file=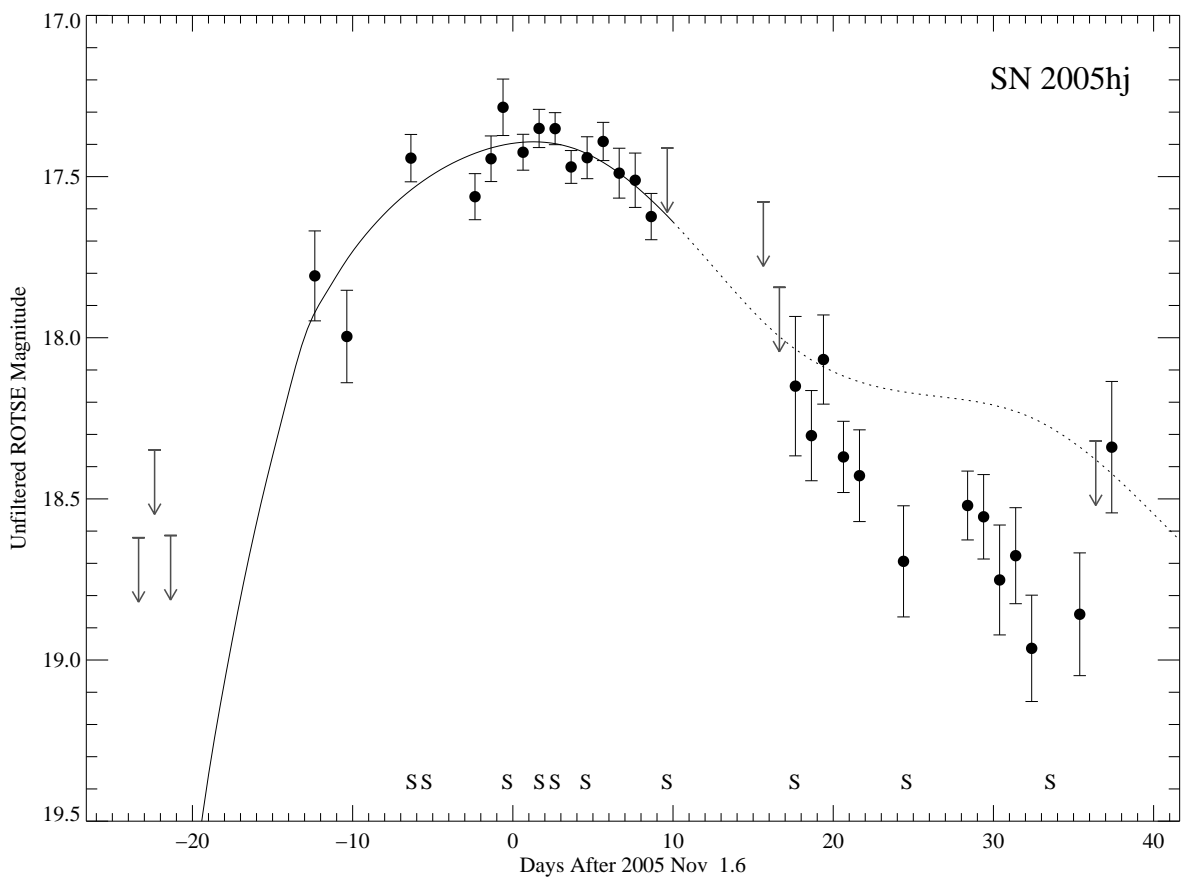,angle=90,width=\columnwidth}
\caption{ROTSE-IIIb unfiltered light curve of SN~2005hj (filled
circles). The best fit $R$-band template from \citet{knop2003} is
plotted as a solid line over the fitting range, and as a dotted line
continuing on to later phases when the rapid decline of the flux below
$\sim$5500~\AA\ causes our unfiltered light curve to fade faster than
the $R$-band decline rate. Arrows mark 5-$\sigma$ upper limits of the
subtractions determined from the noise level in annuli centered on the
location of the SN. Epochs with HET spectra are marked with ``S.''}
\label{lc}
\end{figure}

SN~2005hj was discovered on October 26.13 UT in the field of Abell 194
as part of the TSS. The TSS uses the wide field ($1\fdg85 \times
1\fdg85$) 0.45m ROTSE-IIIb telescope \citep{akerlof03} at the McDonald
Observatory in Texas to scan nearby galaxy clusters nightly for
transients with a modified version of the PSF-matched image
subtraction code from the Supernova Cosmology Project.  SN~2005hj was
found at an unfiltered magnitude (calibrated against the USNO-B1.0 R2)
of $C_R=17.4$ and is located at $\alpha=01^h26^m48\fs27$,
$\delta=-01\arcdeg14\arcmin16\farcs8$.  The foreground reddening at
this location is $E_{B-V}=0.039 ^m$ \citep{schlegel1998}.  Examination
of ROTSE-IIIb images from Oct. 20 and Oct. 22 shows the SN was
detected prior to discovery, but not significantly well to pass the
search pipeline's automatic cuts. Figure \ref{lc} shows the ROTSE-IIIb
light curve for SN~2005hj through 40 days after maximum light. To
construct the light curve, we co-added images taken on a given night
(usually 6) excluding any frames of significantly lower quality due to
passing clouds or wind sheer, and then subtracted the reference image
convolved to the same PSF. Magnitudes were determined by fitting the
local PSF (derived from the co-added nightly images) to the location
of the SN on the subtracted frame using custom software and the
DAOPHOT PSF-fitting routines (\citealt{stetson87} ported to IDL by
\citealt{landsman89}).

The unfiltered CCD response of ROTSE-IIIb has an approximate full
width of $\sim 4000$~\AA\ centered in the $R$-band around 6000~\AA.
Because we do have some sensitivity in the blue and since the $B-V$
colors of SNe~Ia typically grow $\sim 1.0$ mag redder in the 30 days
after maximum \citep{phillips1999,krisciunas2003}, there is a blue
deficit at later times that causes our unfiltered magnitudes to
decline more rapidly than the true $R$-band fading. Note that $V-R$
colors of SNe~Ia are close to zero at maximum light.  We therefore
limit the light curve fitting to data taken before 10 days after
maximum (determined through several iterations of the fit), during
which the color evolution is minimal. The best fit $R$-band template
from \citet{knop2003} is also shown in Figure \ref{lc}. The date of
maximum light determined from the fit is Nov. 1.6 with a formal error
of 0.7 days (note the template phases are relative to the $B$-band
maximum). The best fit stretch factor \citep{perlmutter1997} for the
light curve width is $s=1.2 \pm 0.1$. The preliminary measurement of
the observed $B-V$ color at $V$ maximum from the Carnegie Supernova
Project is $0.07^m \pm 0.05$ after removal of the host light but
before any extinction or $k$-corrections are applied (M. M. Phillips,
private communication).

\begin{figure}
\epsfig{file=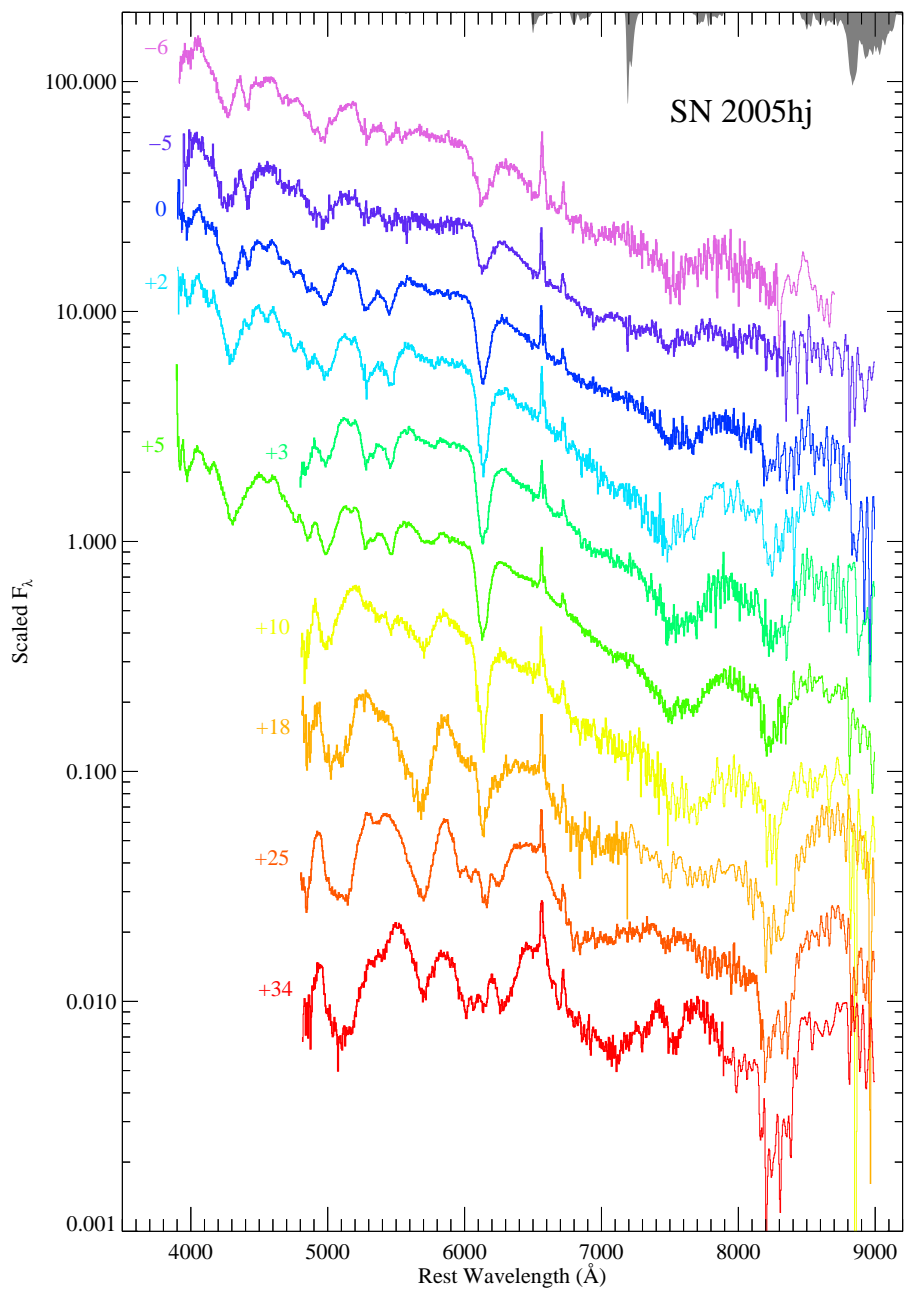,angle=0,width=0.75\columnwidth}
\caption{Spectral evolution of SN~2005hj recorded by the HET from $-6$
to $+34$ days after maximum light (2005 Nov. 1.6). The estimated
galaxy contamination has been subtracted and the spectra have been
shifted for clarity. For display purposes, portions of the spectra
with low signal to noise have been smoothed (thin line segments). The
typical telluric absorption spectrum is shown by the grey shading
along the top of the figure.  }
\label{spec}
\end{figure}

Near real-time photometric analysis combined with target of
opportunity (ToO) time on the neighboring 9.2m Hobby-Eberly Telescope
(HET) allowed us to obtain optical spectra just 4 hours after the
discovery images were taken and every few days over the next 6
weeks. These observations are detailed in Table \ref{spec}.  The
instrumental response is such that very little second order light is
expected blue of 8900~\AA\ even with the GG385 blocking filter.  The
data were reduced in the optimal manner using IRAF\footnote{IRAF is
distributed by the National Optical Astronomy Observatories, which are
operated by the Association of Universities for Research in Astronomy,
Inc., under cooperative agreement with the National Science
Foundation.} and custom IDL scripts. The wavelength scale was
calibrated against Cd and Ne lamps and its accuracy was verified by
comparing night sky lines to the spectral atlas of
\citet{hanuschik2003}. Because the HET pupil size varies for different
tracks, absolute flux calibration cannot reliably be achieved;
however, we used the standard stars of \citet{massey1988} and
\citet{massey_gronwall1990}, which were observed using the same
setups, to achieve {\it relative} spectrophotometric calibration and
to remove strong telluric features.

The redshift of the host galaxy was derived from narrow emission lines
around 7000~\AA\ (observed), which we attribute to H-$\alpha$,
[\ion{N}{2}], and [\ion{S}{2}] in the host galaxy.  We combined all
the spectra and simultaneously fit these lines with Gaussians to
determine the line centers. The line redshifts are best fit by
$z=0.0574 \pm 0.0002$, and we adopt this value for the SN. This gives
SN~2005hj an absolute peak magnitude of $-19.6$ in our unfiltered band
pass (assuming H$_{0}$=71 km s$^{-1}$ Mpc$^{-1}$, $\Omega_{m}=0.3$,
and $\Omega_{\Lambda}=0.7$), and places the host well behind Abell 194
($z=0.0180$; \citealt{struble_rood1999}). The brightness and broad
light curve shape suggest that SN~2005hj is a slightly over-luminous
SN~Ia.

\begin{figure}
\epsfig{file=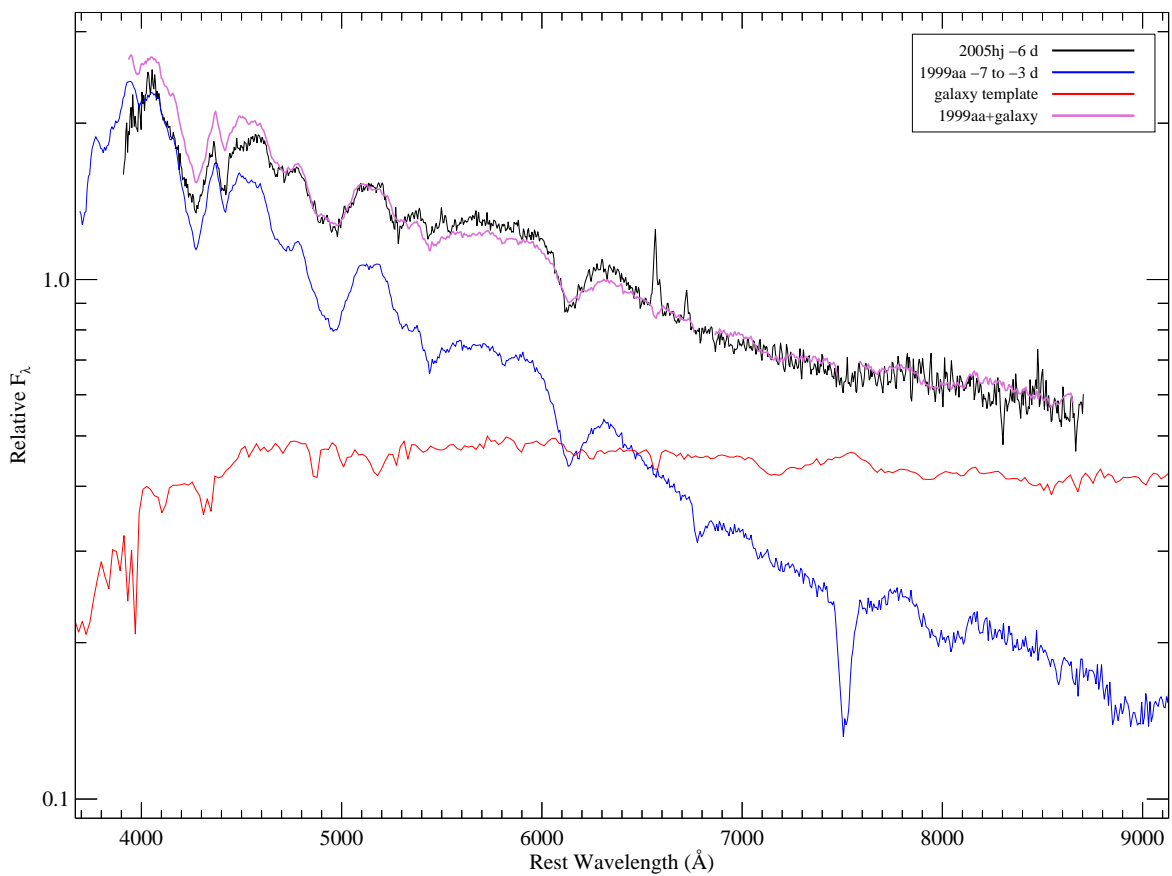,angle=90,width=\columnwidth}
\caption{HET/LRS spectrum of SN~2005hj at $-6$ days (black line). The
blue curve shows the relative contribution from a SN~1999aa-like
supernova at the same phase from a linear interpolation of the $-7$ and
$-3$ day spectra presented by \citet{garavini2004}, and the red line
gives the estimated relative host galaxy light contamination. The
relative scaling for the galaxy and SN components were determined
through a least squares fit to the SN~2005hj spectrum. The purple line
shows the fit (portions of the spectra effected by telluric lines were
excluded from the fitting as shown by gaps in the fit curve).}
\label{spec0}
\end{figure}

The unfiltered ROTSE-IIIb reference image shows that the host for
SN~2005hj is relatively bright ($C_R = 17.8$) and compact, and is
therefore likely a significant contaminant to our spectra. Thus, we
have to subtract the galaxy contribution (see Fig. \ref{spec0}).
Lacking an observed spectrum for the host galaxy excluding the SN
light, we constrained the galaxy SED using archival Sloan Digital Sky
Survey (SDSS) $ugriz$ observations and obtained a template galaxy
spectrum (N. Drory 2005, private communication).  The relative amounts
of SN and galaxy light in the spectral apertures will vary not only
with the changing SN brightness, but also with the seeing, slit width
and positioning. Also plotted in figure \ref{spec0} is a spectrum of
SN~1999aa (blue curve) constructed via a linear interpolation of the
$-7$ day and $-3$ day spectra presented by \citet{garavini2004}.
Noting the similarity of the spectral features of SN~1999aa and
SN~2005hj, we assume that we can model our observed spectra as a
linear combination of our galaxy template and the SN~1999aa spectra
interpolated to the same phase as the SN~2005hj observations. We
perform a least squares fit to determine the relative contributions of
each component. The red line in figure \ref{spec0} shows the derived
contribution of galaxy light in the $-6$ day spectrum. Aside from a
few small differences (most noticeably in the \ion{Mg}{2}
$\lambda$4481 triplet), some of which may be explained by calibration
errors, the combined SN~1999aa + host spectrum (purple curve) is a
good fit. The over all fit is improved if we interpolate the SN~1999aa
spectra to $-5$ days instead of $-6$, especially in the 5400~\AA\ to
6500~\AA\ range, which could imply a $\sim$1 day error in the date of
maximum light or different time scales for the spectral evolutions of
the two SNe. We repeated this process for all the SN~2005hj spectra,
each time using the same galaxy template and the SN1999aa spectra
(interpolated to the appropriate phase) as reference to determine the
relative amount of galaxy light. In general, the galaxy template added
to the SN~1999aa spectra does an excellent job of reproducing the
observed SN~2005hj spectra. The galaxy light typically dominates the
flux red of 7000~\AA. Figure \ref{spec} shows the spectral evolution
of SN~2005hj recorded by the HET between days $-6$ and $+34$ with the
derived galaxy contribution subtracted.

\subsection{Spectral Characteristics of SN~2005hj}\label{spec_char}

\begin{figure}
\epsfig{file=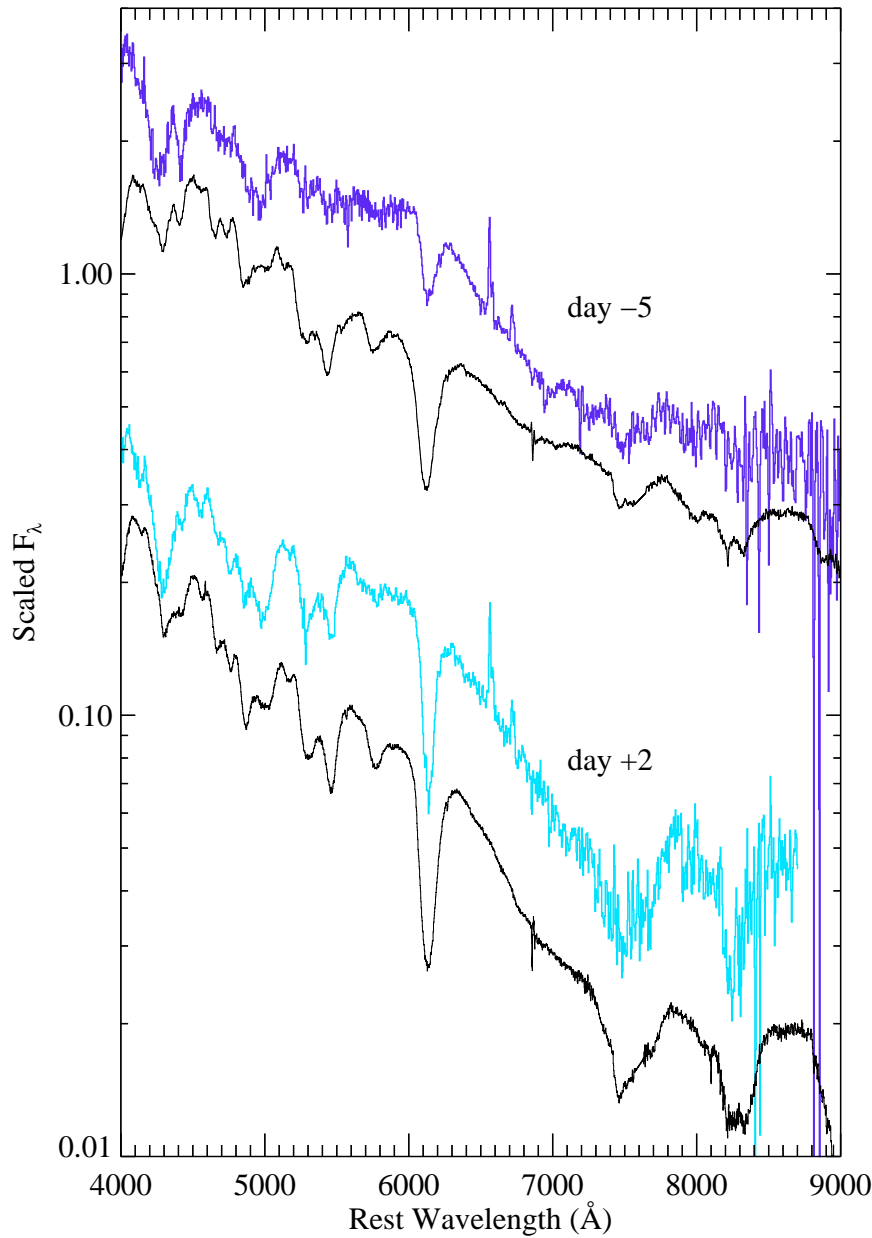,angle=0,width=0.75\columnwidth}

\caption{Comparison of the spectra of SN~2005hj (colored lines) and
SN~1994D (black lines; \citealt{patat1996}) at 5 days before and 2
days after maximum light. The spectra have been shifted for
clarity. At both phases shown, the absorption component of the
\ion{Si}{2} $\lambda$6355 line is more narrow for SN~2005hj. The
relative line depths for this feature are similar at $+2$ days,
however, the \ion{Si}{2} $\lambda$6355 absorption is much weaker for
SN~2005hj in the $-5$ day spectrum. Despite these differences, the
velocities inferred from the minima of the \ion{Si}{2} $\lambda$6355
lines are consistent between SN~2005hj and SN~1994D to within 5\%,
the former being 560 km s$^{-1}$ slower at $-5$ days and 150 km
s$^{-1}$ slower at $+2$ days.}

\label{spec1}
\end{figure}

Overall, SN~2005hj shows spectra with lines dominated by intermediate
mass and iron group elements as is typical for SNe~Ia. While the lines
show normal expansion velocities, the absorption components are more
narrow and, for the early phases, weaker than typically observed, as
exemplified by the \ion{Si}{2} $\lambda$6355 line (see
Fig. \ref{spec1}). SN~2005hj also shows an atypical velocity evolution
of these features over time.

Line minima are useful diagnostic indicators of the ejecta structure
as they give the abundances and velocities of the material.  The
actual measurement of the velocity at the minimum of the line profile
is complicated by the presence of the continuum, other blended lines,
and some uncertainty in the true line profile shape. Detailed modeling
is required to accurately sort out all the components and how they
relate to the photospheric layer to reveal the velocity distribution
of the ejecta\footnote{In general, \ion{Si}{2} lines form above the
photosphere and velocities measured from such absorption minima can be
1000 to 2000 km s$^{-1}$ larger than those measured from weak
lines. However, for shell models the steep density gradients cause
even strong lines to from very close (in radius) to the actual
photosphere.}. Such models have shown that the absorption minima
approximate the photospheric expansion velocities to within about 1000
km s$^{-1}$ at maximum light \citep{hoeflich1995,marion2006}. Thus,
simple line fitting can lead to a rough description of the ejecta
velocities, and allows a useful comparison to discriminate between
different models. At late times the photosphere will recede below the
Si rich layer and so the velocities derived from the \ion{Si}{2}
$\lambda$6355 line will become increasingly discrepant with the
photospheric velocity. For deflagration and classical detonation
models, this departure will begin to set in 1-2 weeks after maximum
light \citep{hoflich_khokhlov1996}.

\begin{figure}
\epsfig{file=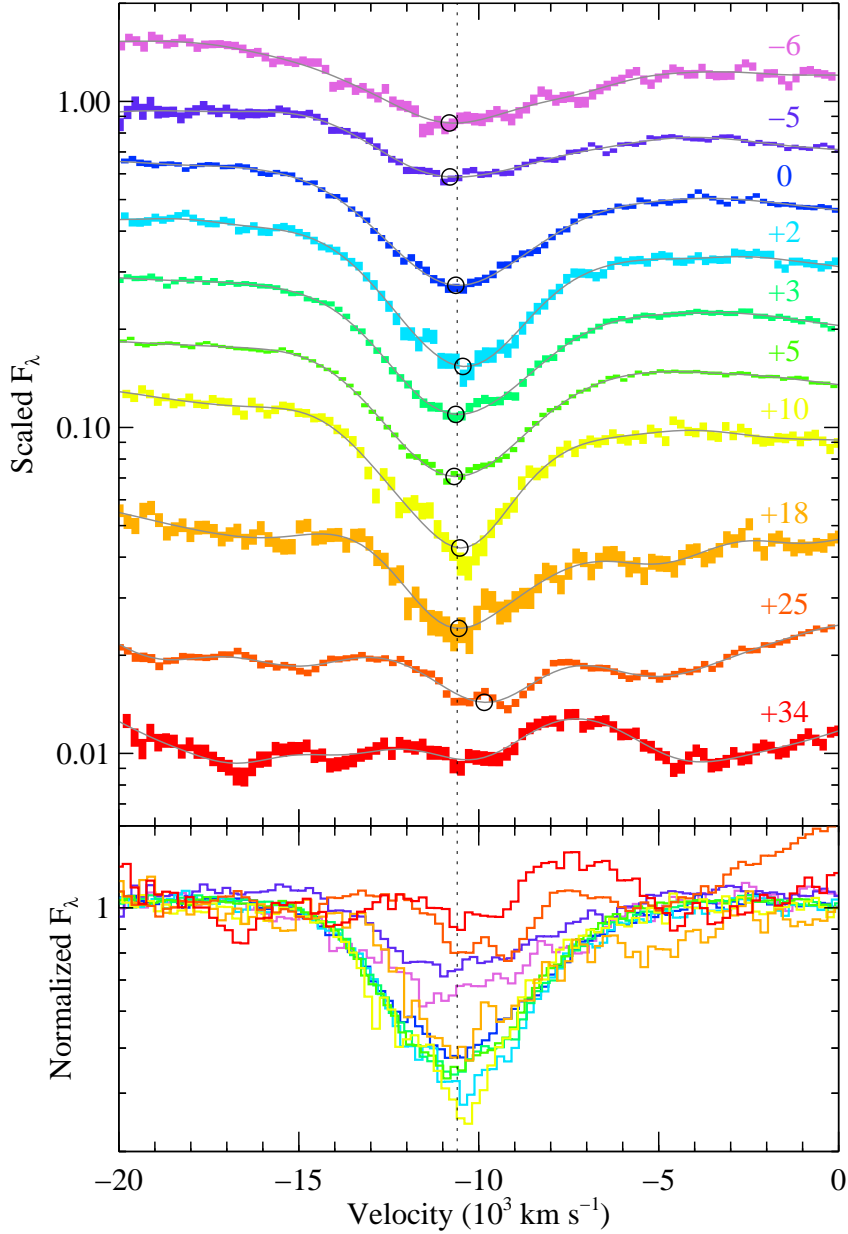,angle=0,width=0.75\columnwidth}

\caption{Evolution of the \ion{Si}{2} $\lambda$6355 line. In the top
  plot, the spectra have been shifted by arbitrary factors for clarity
  with the youngest spectrum on the top and the oldest on the bottom,
  and they are colored as in figure \ref{spec}. Each spectral bin is
  plotted as a rectangle with the height indicating the statistical
  1-$\sigma$ error interval. The effects of the FT smoothing are shown
  by the grey curves, and open circles mark the FT smoothed
  minima. The bottom plot shows the data normalized by the continua to
  emphasize the relative spectral evolution. The vertical dashed line
  marks the plateau velocity of 10,600 km s$^{-1}$.}
\label{SiII}
\end{figure}

The strength of the \ion{Si}{2} $\lambda$6355 line and its persistence
from at least 2 weeks before to 4 weeks after maximum light make it a
valuable tool for probing the ejecta.  Its evolution with time is
shown in figure \ref{SiII} for the case of SN~2005hj. To determine the
velocity, we smooth the spectra by Fourier Transform filtering, divide
by the estimated continuum, and then select the lowest point using
spline interpolation over a selected range. The continuum is
represented by a medium order (6th - 7th) polynomial fit to regions of
the spectra that are not strongly affected by lines. To smooth the
spectra, we use a Fourier Transform to convert the data into a power
spectrum, and then multiply this by a filter to remove high frequency
variations. We then apply a reverse FT to the filtered power spectrum
to recover the smoothed spectrum. The filter has the functional form

\begin{equation}
  \Phi(\nu) = \left\{
  \begin{array}{ll}
    1 & \rm{for}\; -\nu_{c} \le \nu \le \nu_{c}\\
    \exp[-(\nu-\nu_c)^2/2\sigma^2] & \rm{otherwise}
  \end{array}
  \right.
\end{equation}

\noindent The filter cutoff frequency, $\nu_{c}$, and attenuation
scale, $\sigma$, were determined as follows: 1) the spectra were
converted into a power spectrum, $P(\nu)$, via Fourier Transform; 2)
the slope of $\log(P)$ is fit over the noise dominated high
frequencies and interpolated through the low frequencies to determine
the noise spectrum; 3) $\nu_{c}$ is taken as the frequency at which
$\log(P)$ drops to within three times the dispersion about the noise
spectrum; 4) $\sigma$ is chosen such that the slope of
$\log[P(\nu_{c}+2\sigma)]$ is twice the noise spectrum slope
(i.e.. $\nu=\nu_{c}+2\sigma$ is the frequency above which noise is
clearly the dominate component). For this analysis, only the spectral
bins with signal to noise above 25 were considered (note the peak
throughput for HET/LRS is near the \ion{Si}{2} $\lambda$6355
line). For consistency, we adopt a single filter for all our analysis,
choosing the results from our nosiest data,
$\nu_{c}=0.0066$~\AA$^{-1}$ and $\sigma=0.0053$~\AA$^{-1}$, which
removes noise in the data but also some real information related to
``sharp'' features in the spectra such as the narrow core to the
\ion{Si}{2} $\lambda$6355 absorption in the day $+10$ spectrum.

\begin{figure}
\epsfig{file=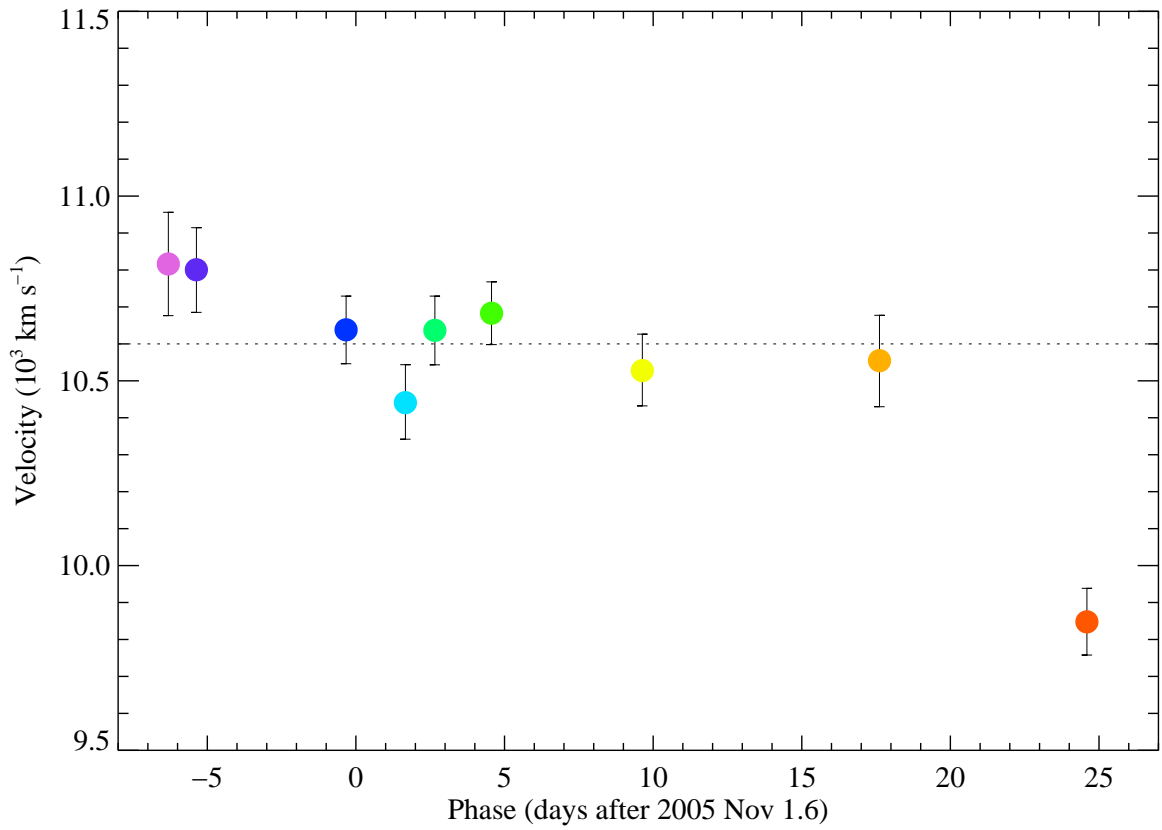,angle=0,width=\columnwidth}
\caption{Velocity of the \ion{Si}{2} $\lambda$6355 line minimum as
determined from the FT smoothed minima. The error bars are the
quadrature sum of the 1-$\sigma$ confidence intervals found via Monte
Carlo simulations and the 80 km s$^{-1}$ scatter found in the \ion{H}{2}
region line centers among the different epochs.}

\label{linevel}
\end{figure}

Using the relativistic Doppler formula and the $gf$-weighted
\ion{Si}{2} $\lambda$6355 rest velocity in the host galaxy frame, we
convert the wavelengths of the line profile minima into expansion
velocities. For each spectrum we conducted 250,000 Monte Carlo
simulations in which normally distributed noise based on the
statistical flux errors was added to the data and the FT smoothed
minimum was found. The peak of the distribution and the interval
containing 68\% of the simulation results were used to calculate the
velocity of the minimum and its error, respectively. We also measured
the relative shift in the \ion{H}{2} region lines over all epochs and
found the scatter to be 80 km s$^{-1}$, which we add in quadrature to
the individual errors. The results are given in table \ref{linedata}
and plotted in figure \ref{linevel}. We find that the data points are
at 10,600 $\pm$ 150 km s$^{-1}$ between maximum light and $+18$ days,
somewhat faster prior to maximum, and significantly slower on day
$+25$. By day $+34$, the \ion{Si}{2} $\lambda$6355 absorption has all
but completely disappeared.

From maximum light through day $+10$, the \ion{Si}{2} $\lambda$6355
line profile shows little change in both depth and width in addition
to maintaining a constant absorption minimum velocity. Of specific
relevance is the blue wing of the absorption profile; this section of
the line is formed by the material at the greatest distance from the
photosphere and at the highest velocities, and as such it should be
the first to vanish as the photosphere recedes. The consistency of
this blue wing from maximum light through day $+10$ suggests the
photosphere falls within the Si enriched layers for at least this
period. By day $+18$ the blue wing has shifted significantly to the
red, while the red wing remains constant except for the effects of an
\ion{Fe}{2} blend around 6250~\AA. Other \ion{Fe}{2} features begin to
appear or strengthen at this phase as well. This behavior could be a
signal that the Si layers are becoming detached from the photosphere
by day $+18$.

\begin{figure}
\epsfig{file=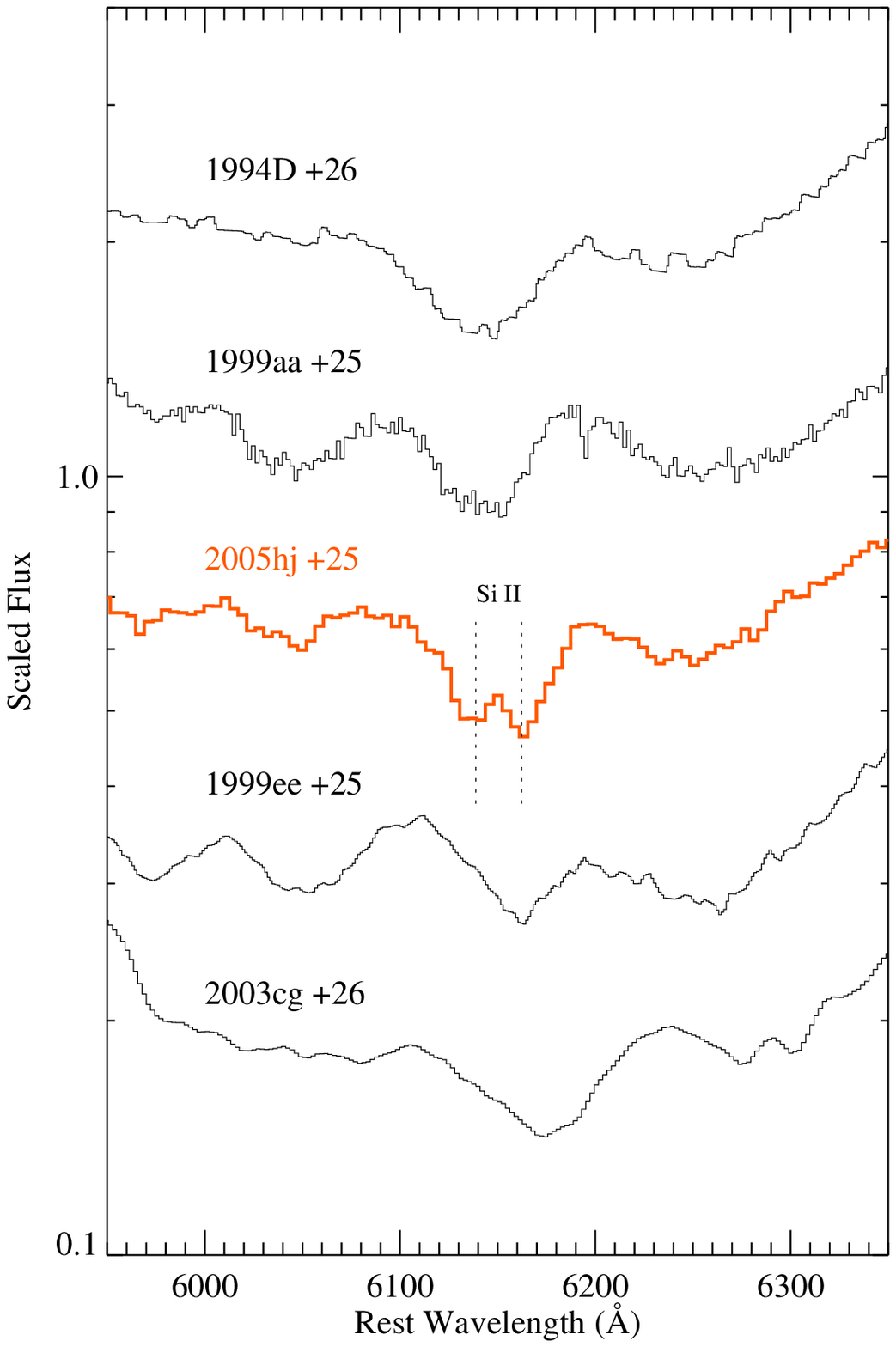,angle=0,width=0.75\columnwidth}
\caption{Spectra near the \ion{Si}{2} $\lambda$6355 doublet for
 SN~2005hj and other SNe~Ia around 25 days after maximum light (1994D;
 \citealt{patat1996}; 1999aa; \citealt{garavini2004}; 1999ee;
 \citealt{hamuy2002}; 2003cg; \citealt{elias-rosa2006}). Vertical
 dotted lines mark the doublet components blue shifted by 10,000 km
 s$^{-1}$. The doublet is clearly resolved only in the case of
 SN~2005hj, while the flat bottom to the SN~1999aa profile may suggest
 similar behavior. }

\label{day25}
\end{figure}

The day $+25$ spectra show a double minimum at the location of the
\ion{Si}{2} $\lambda$6355 feature (see figure \ref{day25}). Telluric
absorption is weak in this wavelength range, and the line profile is
clearly seen in each of the three individual exposures, which support
the reality of this feature.  A possible explanation for this feature
is contamination from the host that is not removed by the template
subtraction; however, galaxy spectra do not typically exhibit features
in this range that could cause such interference, and even if such
were the case, we would expect to see similar behavior in the $+34$
day spectra. A second possibility is contamination from \ion{Fe}{2}
lines. Using the spectral analysis tool SYNOW
\citep{jeffery_branch1990,fisher1997,fisher1999}, and the example of
SN~1994D as a starting place \citep{branch2005}, we find that while
\ion{Fe}{2} likely produces the absorption dips $\sim 100$~\AA\ away
on either side of the \ion{Si}{2} $\lambda$6355 line, it is unlikely
responsible for the double minimum. The third possibility, which we
favor, is that this double minimum simply appears because we are
resolving the \ion{Si}{2} $\lambda$6355 doublet. This result implies
that the \ion{Si}{2} seen in the $+25$ day spectra is confined to a
very narrow region of velocity space ($\Delta v \approx 1,500$ km
s$^{-1}$). If accurate, the true minimum of the \ion{Si}{2}
$\lambda$6355 doublet would be about 100-200 km s$^{-1}$ faster than
indicated in figure \ref{linevel} and table \ref{linedata}, but still
significantly below the plateau velocity. The emergence of this thin
layer may also be responsible for the appearance of the narrow core in
the $+10$ day spectrum as well as the apparent double minimum to the
$+18$ day data.  Some remnant of the blue component to the doublet may
persist to the $+34$ day spectrum. Figure \ref{day25} also shows the
spectra of several other SNe~Ia taken around 25 days after maximum
light. While the distinctly double minimum appears unique to
SN~2005hj, the width and depth of the \ion{Si}{2} feature is roughly
consistent with the others.
 
SN~2005hj clearly belongs to the low velocity gradient (LVG) group in
the classification scheme of \citet{benetti2005}, but moreover the
velocity derivative from maximum light through day $+18$, $\dot{v}=3
\pm 7$ km s$^{-1}$ day$^{-1}$, is consistent with no
change\footnote{As defined by \citet{benetti2005}, $\dot{v}$ is the
average daily rate of decrease in the expansion velocity from maximum
light through the last available spectrum before the \ion{Si}{2}
$\lambda$6355 line disappears; therefore, including the day $+25$
spectrum, SN~2005hj formally has $\dot{v}=27 \pm 4$ km s$^{-1}$
day$^{-1}$, but with a $\chi^2$ per degree of freedom of 3.2}. From
the line profile evolution (table \ref{linedata}, figures \ref{SiII}
and \ref{linevel}) we can deduce a plateau phase starting at $-2.5 \pm
2.5$ days which lasts no more than 30 days. Noting the change in the
\ion{Si}{2} $\lambda$6355 line profile in the $+18$ day spectrum, we
conservatively mark the end of the plateau phase as day $17.5 \pm 7.5$
days, which gives the plateau phase a total duration of $20 \pm 10$
days.
The \ion{Si}{2} $\lambda$6355 velocity evolution derived from the
minima of FT smoothed spectra of several selected SNe~Ia is plotted in
figure \ref{linevels}. The velocity plateau of SN~2005hj is similar to
that of other over-luminous SNe~Ia such as
SN~1999aa~\citep{garavini2004} and SN~2000cx~\citep{li2001}, but it is
distinct from normal SNe~Ia such as SN~1994D~\citep{patat1996} and
SN~1992A~\citep{kirshner1993} that do not show a plateau
phase\footnote{We removed the data point at day $-3$ from the
SN~1999aa curve because all spectral features in these data seem to be
frequency shifted including the telluric features.}.

\section{Physical Constraints from Explosion Models}\label{models}
 
There is general agreement that SNe~Ia result from some process
involving the combustion of a degenerate C/O white dwarf
\citep{hf60}. Within this general picture, two classes of models are
most likely. The first is an explosion of a C/O-WD with a mass close
to the Chandrasekhar limit ($M_{Ch}$) that accretes matter through
Roche-lobe overflow from an evolved companion star
\citep{Whelan73}. In this case, the explosion is triggered by
compressional heating near the WD center.  Alternatively, the SN could
be an explosion of a rotating configuration formed from the merging of
two low-mass WDs, after the loss of angular momentum
\citep{it84,pac85}. Candidate progenitor systems have been observed
for both scenarios: WD binary systems with the correct period to merge
in an appropriate time scale with an appropriate total mass
\citep{max2000}; and supersoft X-ray sources
\citep{greiner91,vdh92,rap94,kah97} showing accretion onto the WD from
an evolved companion. There are still open questions about the details
of both the merging and accretion processes
\citep[e.g.][]{nomoto82,benz90,piersanti2003,nomoto03}.

From the observed spectral and light curve properties, the first
scenario appears to be the most likely candidate for the majority of
normal SNe~Ia.  In particular, delayed detonation (DD) models
\citep{khokhlov1991,yamaoka92,woosley_weaver1994} have been found to
reproduce the majority of the observed optical/infrared light curves
(LC) and spectra of SNe~Ia reasonably well
\citep{hoeflich1995b,fisher95,hoflich_khokhlov1996,lentz2001,hoflich2002,marion2003,marion2006}. In
the DD scenario, a slow deflagration front turns into a
detonation. The WD pre-expands during the deflagration and undergoes
complete burning during the detonation phase. Similarly, the classical
deflagration models W7 \citep{nomoto1984} show similar behavior to DDs
but only by neglecting instabilities due to the deflagration fronts
\citep{gamezo2003}.  For recent reviews see
\citet{branch1998,hillebrandt2000,hoeflich2006b}.

\begin{figure}
\epsfig{file=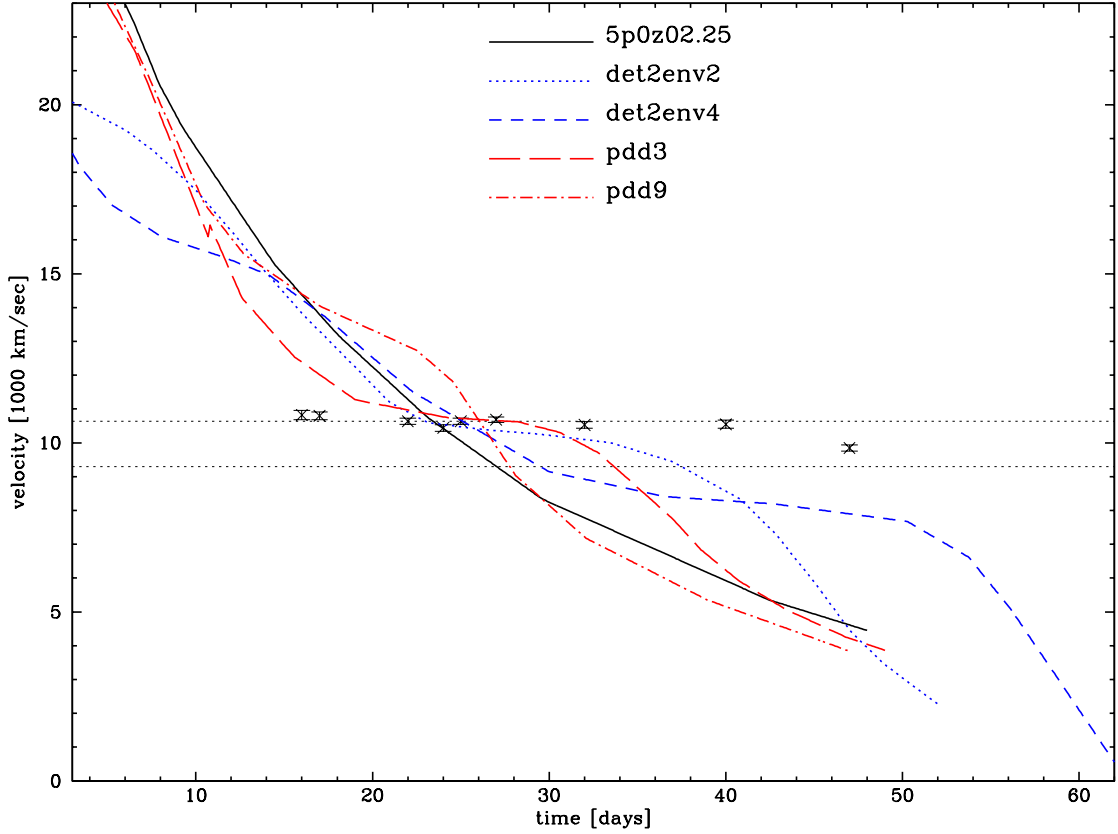,angle=270,width=\columnwidth}

\caption{Photospheric velocity versus time since explosion for models
of Branch-normal SNe~Ia. The photosphere is defined by the radius of
last scattering, and because Thomson scattering is the dominate source
of continuum opacity, this position is largely wavelength
independent. Shown are the classical delayed detonation model {\tt
5p0z02.25} \citep{hoflich2002}, the pulsating delayed detonation
models {\tt pdd3} and {\tt pdd9} \citep{khokhlov1993,hoflich1995}, and
the tamped detonation models {\tt det2en2}/{\tt 4}
\citep{hoflich_khokhlov1996}. The X's mark the velocities of SN~2005hj
as measured from the minima of the \ion{Si}{2} $\lambda$6355 line
assuming a 22 day rise to peak. The dotted horizontal lines mark the
region with $>1$\% Si abundance in model {\tt pdd3}. As discussed in
the text, the \ion{Si}{2} $\lambda$6355 line minimum is an imperfect
indicator of the photospheric velocities, and a systematic offset can
develop at later times once the photosphere has receded below the Si
rich layers. This may explain the discrepancy between the model
predictions and the last two observations, which show weakening Si and
strengthening Fe lines. The low velocities observed at early times for
SN~2005hj may indicate that rather little outward mixing took place.
}

\label{modelvel}
\end{figure}

Despite the success of classical DD and W7 models, both lack the basic
features seen in SN~2005hj.  Neither predicts a long plateau in
velocity; they instead show a smooth decline of the photospheric
velocity\footnote{Although the \ion{Si}{2} $\lambda$6355 line is an
imperfect tracer of the photospheric velocity as mentioned in
\S\ref{spec_char}, the observed SNe~Ia population typically exhibits a
1000-3000~km~s$^{-1}$ decrease in the measured line velocities between
1 week before maximum light to 2 weeks
after\citep{branch1988,benetti2005}, and the deflagration and
classical delayed detonation models employed to explain these events
have shown a correspondingly large decrease in photospheric velocities
over the same period \citep{khokhlov1993}. These models are
inconsistent with the corresponding $\approx 300$~km~s$^{-1}$ shift
measured for SN~2005hj.} as a function of time
(Fig. \ref{modelvel}). This happens because in expanding envelopes the
photosphere recedes in mass and, because of the homologous expansion,
in velocity as well.  This behavior results from the smoothly
declining density structure of the WD and the fact that variations in
the specific energy production are small.
 
In contrast, shell-like density structures will produce velocity
plateaus in a natural way because the photosphere remains in the shell
for some time as shown by \citet{khokhlov1993,hoflich_khokhlov1996}.
To form a shell-like structure requires interaction of rapidly
expanding material with a surrounding envelope. Various mechanisms
have been suggested to supply this surrounding matter: the pulsating
delayed detonation scenario \citep{hoeflich1996}, mergers or tamped
detonation models.  Shells may also form by the interaction of an
exploding single WD within the progenitor system
\citep{gerardy2004,quimby2006}.

We analyzed the observations of SN~2005hj based on detailed, spherical
models for supernovae published in the literature. The models are
based on detailed calculations for the explosion, light curve and
spectra.  The models considered include delayed detonations,
deflagrations, pulsating delayed detonations and tamped
detonation/merger scenarios.  In figure \ref{modelvel}, we show the
photospheric velocities as a function of time for these models along
with the Branch-normal SNe~Ia to illustrate the formation of a plateau
in the models that naturally form a shell. Note for lower shell
masses, this ``plateau'' is more accurately described as a period of
slowly declining velocities.

In classical delayed detonation models and for normal-bright SNe~Ia,
Si is present over a wide range in mass, spanning about $ 0.4 $ to
$0.5 M_\odot$, which corresponds to velocities from about 8,000 to
9,000 km s$^{-1}$ to more than 20,000 km s$^{-1}$.  The Si layer is
thick (in the mass frame) because explosive oxygen burning occurs over
a wide range of temperatures. The density gradient is smooth and Si is
mostly in \ion{Si}{2}, so initially the velocity derived from the
minimum of the \ion{Si}{2} $\lambda$6355 line smoothly declines with
the receding photosphere governed by the geometrical dilution of the
expanding envelope. Eventually, the photosphere begins to recede below
the Si layer at which point the evolution of the \ion{Si}{2}
$\lambda$6355 line profile will show the following behavior: 1) the
optical depth of the highest velocity material at the largest radii
will begin to decline below 1 and as such the blue wing of the line
profile will start to drift toward the red; 2) as the optical depth
decreases, the strength of the line as measured from the line depth
will decrease; 3) the line minimum may continue to slow, but it will
grow increasingly discrepant with the photospheric velocity. This
phase typically begins 1-2 weeks after maximum light for normal SNe~Ia
and is heralded by the appearance of \ion{Fe}{2} lines. While this
behavior is commensurate with observations of normal-bright SNe~Ia
such as SN~1994D, this behavior is not consistent with the
observations of SN~2005hj.

The \ion{Si}{2} $\lambda$6355 line seen in SN~2005hj is narrow, and
during the plateau phase the wings do not change, the depth does not
change, and the velocity of the minimum does not change to within the
errors. The data require a narrow distribution of \ion{Si}{2} in
velocity space, and we suggest this may be explained by an interaction
that compresses the Si rich layers as predicted by merger and
pulsating delayed detonation models. The shell models are also
consistent with the velocity drop seen after the plateau because a
significant amount of Si is located below the shell
\citep{khokhlov1993,hoflich_khokhlov1996}.

In Fig. \ref{modelprop}, we show general properties of these models.
As discussed in the papers above, to first order, the observational
signatures of the shell depend on the mass of the shell $M_{shell}$.
Almost the entire WD is burned, and momentum conservation determines
the amount of high velocity matter that can pile up in the expanding
shell.  With increasing shell mass, more material of the SN envelope
is slowed down. As a consequence, the velocity of the shell
$v_{shell}$ decreases with $M_{shell}$. Because it will take longer
for the photosphere to recede into the lower velocity matter, the time
until the beginning of the plateau phase, $t_0$, increases with
$M_{shell}$.  The optical depth increases with $M_{shell}$, duration
of the plateau, $\Delta t_{shell}$, also increases, the temperature
gradient becomes steeper, and the photosphere becomes cooler
(i.e. $B-V$ increases) with increasing $M_{shell}$
\citep{khokhlov1993,hoflich_khokhlov1996}.  The duration of the
plateau, $\Delta t_{shell}$, is defined by the velocity spread $\delta
v $ around $v_{Shell}$ with $\delta v = 500$ km s$^{-1}$, which puts
the end of the plateau phase safely into the parts of a rapidly
declining $v_{ph}$.  We choose a larger value than in the observations
to avoid ambivalences due to discreetness, which, in some of the
models, is of the order of $\approx 100$ km s$^{-1}$. By increasing
$\delta v $ from 200 to 500 km s$^{-1}$ the nominal duration is
increased by $\approx 1$ day. However, we also note that the actual
width depends on the velocity spread in the shell (see \S
\ref{conclusions}).

Given the model predictions, we can use different observational
indicators to test which $M_{shell}$ is consistent with SN~2005hj
(Fig. \ref{modelprop}). All three parameters, $v_{shell}$,
$t_{shell}$, and $t_0$ suggest $M_{shell} \approx 0.2 M_\odot$, with
the allowed ranges specifically bracketed by 0.15-0.6, 0.1-0.25, and
0.1-0.25 $M_\odot$ for the plateau length, shell velocity, and plateau
onset, respectively, taking the observed errors into account. The
comparison between the $B-V$ color as a function of $v_{shell}$,
$t_{shell}$, or $t_0$, however, shows only marginal consistency
between the observations and the models if we assume only foreground
redding by the Galaxy. We note, that the intrinsic $B-V$ color of the
models is uncertain by about 0.05 to $0.1^m$ at maximum light.  The
two best-fitting models, {\tt pdd3} and {\tt det2env2}, show a peak
brightness, $M_{V}$, of $ -19.42^m$ and $-19.41^m$, respectively, with
an uncertainty of $\approx 0.1 ^m$ \citep{hoflich_khokhlov1996} vs. a
typical DD model with $-19.2^m$ \citep{hoflich2002}, i.e.  they are
brighter by about 20 \% mostly due lower escape probablity of
$\gamma$-rays that results when the $^{56}$Ni layers are slowed down
because of the interaction \citep{hoflich1991}.

\section{Discussion and Conclusions}\label{conclusions}

We have presented photometric and spectroscopic data for SN~2005hj, a
slightly over-luminous Type Ia. The most striking feature is an
apparent plateau in the expansion velocity evolution, which we derive
from the \ion{Si}{2} $\lambda$6355 line. The velocities remain at
about $10,600$ km s$^{-1}$ for about 3 weeks starting slightly before
maximum light, and this plateau is bracketed by preceding and
succeeding decelerations. We find that Si is confined to a relatively
narrow velocity region. Analysis of the detailed observations in
concert with published models suggest there may be some physical
distinction between SN~2005hj and other normal-bright SNe~Ia that may
systematically affect their use as distance indicators if not properly
taken into account.

The models considered include delayed detonations, deflagrations,
pulsating delayed detonations and tamped detonation/merger scenarios.
In order to explain the narrow \ion{Si}{2} $\lambda$6355 line and its
plateau in velocity, we suggest an early interaction that forms a
dense shell as predicted by merger and PDD models. The spectral and
photometric peculiarities are consistent with respect to the velocity,
duration, and onset of the plateau, and marginally consistent with the
maximum light color, for models that have shells of about $0.2
M_\odot$. As indicated by earlier works
\citep{khokhlov1993,hoflich_khokhlov1996}, the mass of the interacting
shell has been found to be the parameter that dominates the details of
these observational signatures independent of how this shell may
form. The tight predicted relation between each of $v_{shell}$,
$t_{shell}$, and $t_0$ may provide a stable means to separate
SN~2005hj like events from regular Branch-normal SNe~Ia. Although the
agreement between the shell models and the observations is good, the
predictions are not necessarily unique and other possibilities may
exist. For example, we have not considered 3-D models such as the
detonation from a failed deflagration scenario recently examined by
\citet{kasen_plewa2006}. For SN~2005hj then the agreement of the
plateau velocity and its duration to that predicted by shell models
may simply be a fluke, and in such case this concordance should then
not hold for other SNe with similar \ion{Si}{2} $\lambda$6355
evolution. Given the data and models considered, we suggest either
PDDs or merger events are responsible for SN~2005hj, and this implies
the existence of two different progenitor channels.

It is important to understand how these two progenitor channels, which
may occur in relatively varying fractions as a function of redshift,
will impact studies using SNe~Ia as distance indicators.
\citet{li2000} estimate that 20\% of SNe~Ia in their sample are either
1991T-like or 1999aa-like. These SNe show spectral features and a
velocity plateau similar to SN~2005hj. \citet{branch2001} found 5
1999aa-like events in the \citet{li2000} sample out of 20 total SNe~Ia
that were observed early enough to show 1999aa-like spectral features,
and one that was 1991T-like; however, in the pre-LOSS sample they do
not classify any of the 7 SNe~Ia with early spectra as
1999aa-like. These nearby samples are constructed from targeted galaxy
searches that have different selection biases than the deep
cosmological surveys, but we will assume a uniform 2005hj-like rate of
25\% for all SNe~Ia. SNe~Ia that appear spectroscopically similar to
SN~2005hj in a single epoch could none the less arise from different
progenitors, and the mass of the low-density envelope around PDDs or
mergers may effect their peak magnitudes and/or light curve shapes,
but we will further assume that all such events deviate uniformly from
the LWR of Branch normal SNe~Ia. \citet{hoeflich1996} calculated the
relation between peak $V$ band magnitudes, $M_V$, and the fading
between maximum light and +20 days, $dM_V(20)$, for a variety of
theoretical models and found that shell models produced $dM_V(20)$
that were 0.2 to 0.3 mag smaller than for (in vacuum) delayed
detonations reaching the same peak magnitude. Therefore using the same
LWR for shell models will result in corrected peak magnitudes
systematically offset by 0.1 to 0.2 mag. Also, the observed peak
magnitudes of SNe~Ia are usually corrected for absorption along the
line of sight using the observed $B-V$ color at maximum light and a
reddening law. For events that are intrinsically red, this will
increase the estimated peak magnitude above its already over-luminous
intrinsic value. Cosmological studies may therefore need to remove or
at least separately handle SN~2005hj-like events to avoid systematic
errors in distance.

\begin{figure}
\epsfig{file=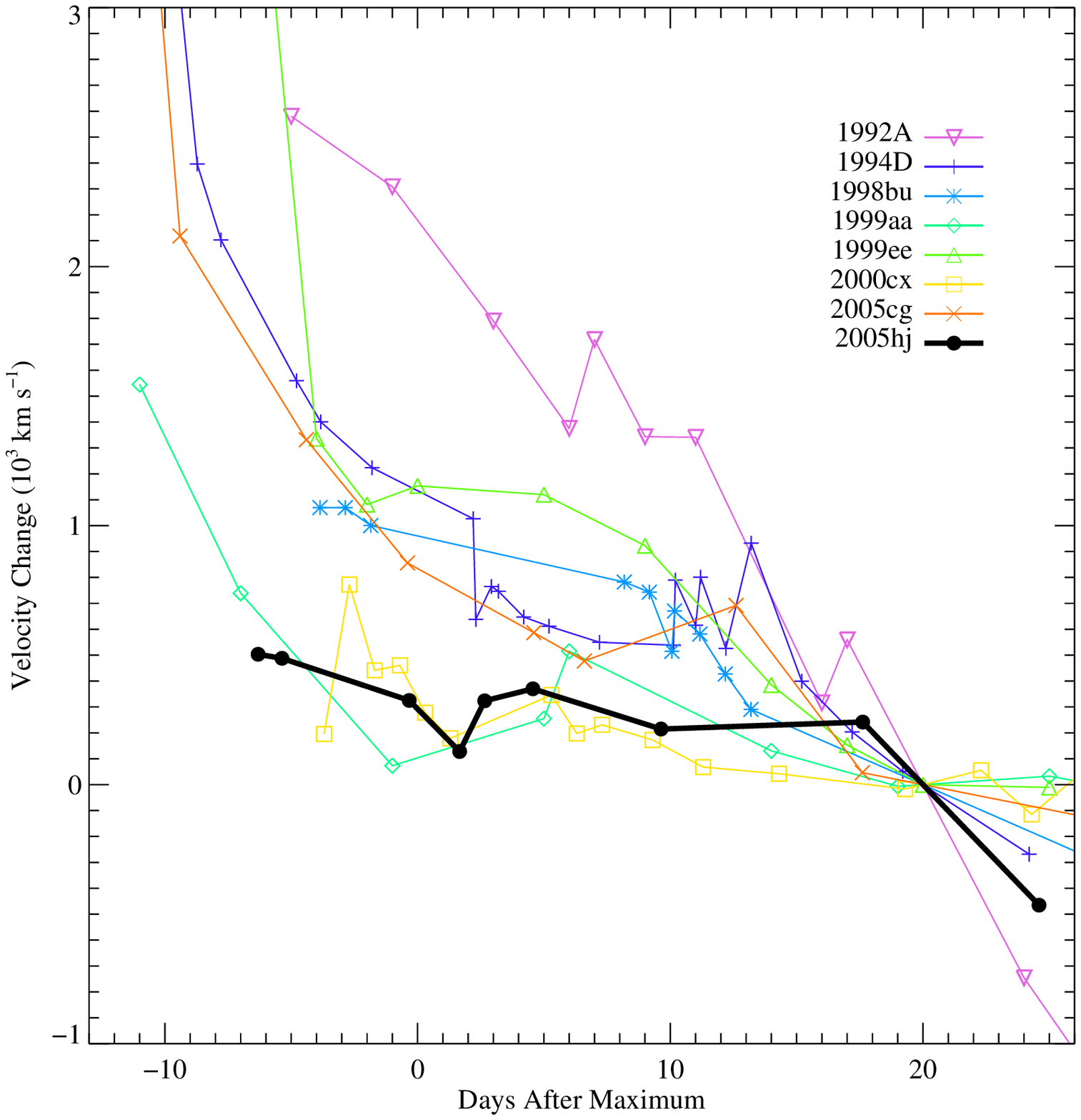,angle=0,width=\columnwidth}
\caption{Velocity of the \ion{Si}{2} $\lambda$6355 line minima for
SN~2005hj and other well-observed SNe~Ia (1992A;
\citealt{kirshner1993}; 1994D; \citealt{patat1996}; 1998bu;
\citealt{jha1999}; 2000cx; \citealt{li2001}; 1999aa;
\citealt{garavini2004}; 1999ee; \citealt{hamuy2002}; 2005cg;
\citealt{quimby2006}).  All velocities were calculated from the FT
smoothed minima using the same filter.  The ordinate shows the change
in velocity relative to the (interpolated) day $+20$ value. Lines
connecting the data points are intended only to guide the eye. The
sample was not drawn randomly from the parent population; instead it
was tailored to show the range of behaviors present in well-observed
in normal-bright events.}

\label{linevels}
\end{figure}

\begin{figure}
\epsfig{file=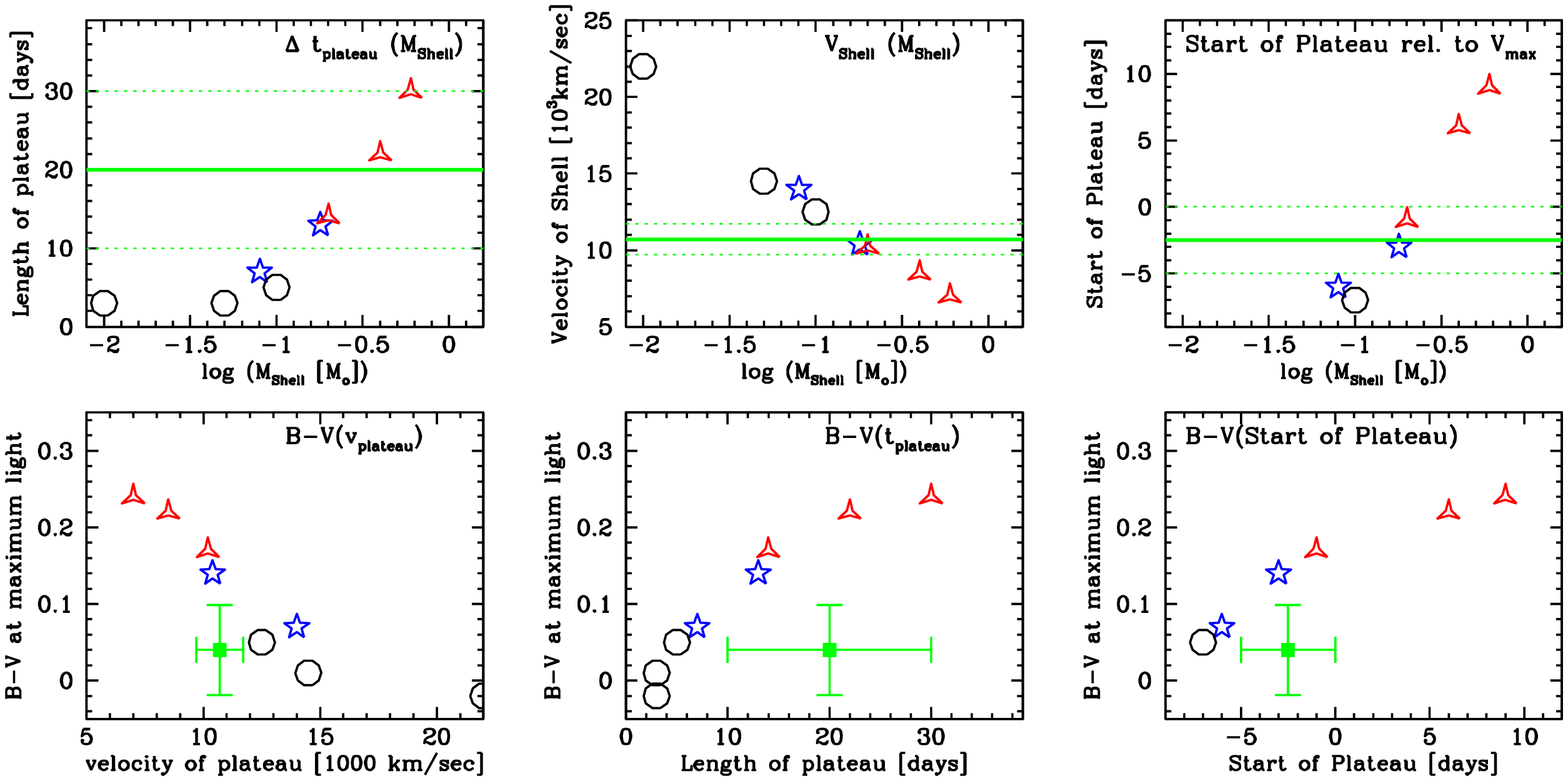,angle=270,width=\columnwidth}
\caption{ Properties of Branch-normal SNe~Ia models with interacting
  shells of mass $M_{Shell}$ including delayed detonations ({\tt
  5p0z22.25} with $M_{Shell}=0.01, 0.05$ and $0.1 M_\odot$, open
  circles \citealt{gerardy2004}), pulsating delayed detonation models
  (open stars; {\tt pdd3};
  \citealt{khokhlov1993,hoflich_khokhlov1996}; {\tt pdd9};
  \citealt{hoflich1995}) and merger models (open triangles; {\tt
  det2env2/4/6}; \citealt{khokhlov1993,hoflich_khokhlov1996}).  In the
  upper plots, we give as a function of $M_{Shell}$ the duration of
  the velocity plateau $\Delta t_{plateau}$, its velocity
  $v_{plateau}$ defined by $\delta v \leq 500$ km s$^{-1}$ (see text),
  and its starting point $t_{0}$ relative to maximum light. In the
  lower plots we give observational relations, namely $B-V$ at maximum
  light as a function of $v_{plateau}$, $\Delta t_{plateau}$ and
  $t_0$. Note that $B-V$ is uncertain by about $0.05$ to $0.1^m$. In
  addition, we give the observations of SN~2005hj (filled squares)
  including error bars.  }
\label{modelprop}
\end{figure}

As a case for the importance of separating different progenitors, let
us consider SN~1999ee. Very similar to SN~2005hj, SN~1999ee shows a
plateau with $v_{shell}=10,500$ km sec$^{-1}$, a duration of $14\pm 3
$ days, and an onset at day $-3 \pm 1 $ relative to maximum
(\citealt{hamuy2002}; see Fig.  \ref{linevels}).  The $B-V$ color of
SN~1999ee was also quite red at maximum light; $+0.28^m \pm 0.04$
after correction for galactic extinction
\citep{stritzinger2002,krisciunas2004}. Based on the standard
brightness decline relation and the corresponding colors,
\citet{hamuy2002} derived reddening in the host galaxy of $0.28 \pm
0.04$ which implies an absolute brightness of $M_V=-19.95^m$ similar
to SNLS-03D3bb which \citet{howell2006} attributed to a
super-Chandrasekhar mass WD. Taking into account the spectroscopic
information about the velocity plateau, its length and onset, we
attribute a portion of the red color to the intrinsic properties of
the supernova. We find that the duration of the velocity plateau, its
onset and size are consistent with a shell mass of $0.2 M_\odot$ which
suggests an intrinsic color $B-V$ of $0.15^m \pm 0.02$ (see
Fig. \ref{modelprop}).  This reduces the reddening in the host galaxy
to $\approx 0.13^m$ and the absolute brightness $M_V$ to $-19.53^m$,
which compares favorably to the model predictions of $-19.42^m$ and
$-19.41^m$ for {\tt pdd3} and {\tt detenv2}, respectively, within the
model uncertainties. Note that there is an interstellar sodium line in
the spectra that implies some degree of reddening within the
host. There are some apparent spectral differences when compared to
SN~2005hj, namely SN~1999ee has a slightly broader blue wing in
\ion{Si}{2} and stronger absorption around 4900~\AA. This may either
hint toward different explosions scenarios (i.e. pulsations versus
mergers), or different viewing angles of asymmetric envelopes.

This brings us to the limitation of our studies. Except for the color,
SN~2005hj fits remarkably well with the merger and PDD model
predictions but, still, it is a single event and the good agreement
may be coincidental. We need a large, well-observed sample of similar
objects to test and verify or falsify the models and to determine the
shell mass distribution. Moreover, 3D effects have been neglected. In
reality, we must expect some dispersion. Though pulsating delayed
detonation models may be expected to be rather spherical, mergers may
be asymmetric with directionally dependent luminosities and colors. In
fact, both classes may be realized in nature.  As mentioned above, the
duration of the plateau, $\Delta t_{shell}$, is defined by the
velocity spread around $v_{Shell}$. The physical width of the shell
depends, to first order, on the distance at which the interaction
occurs and the density distribution of the interacting expanding media
and shell during the hydrodynamical phase of the interaction
\citep{gerardy2004}. For obvious reasons, asymmetries of the shell
will increase the velocity gradient seen over the shell. The
observations of SN~2005hj indicate a very flat plateau that, in
principle, may further constrain the properties of the shell. For
SN~2005hj, this may already indicate a rather spherical shell and hint
toward the PDD scenario or mergers with an intermediate disk of very
large scale heights.  However, additional information needs to be
taken into account such as detailed line profiles and statistical
properties to break this degeneracy between mergers and PDDs.

As a next step, detailed models for the explosion, light curves and
spectra tailored toward individual events need to be
constructed. Whereas the mean velocity of the shell for a given mass
is dictated by momentum conservation, the thickness of the shell is
limited by the distance of the shell material, the distance sound can
travel during the interaction, and the specific density profile within
the shell. With increasing distance of the shell, the relative size
(and corresponding velocity spread) becomes smaller because the sound
speed remains about the same. The intrinsic color will be sensitive to
the optical depth of the shell, which is governed by the magnitude of
the density jump and thus depends on the distance of the interacting
shell from the WD \citep{gerardy2004}. The blue $B-V$ color for
SN~2005hj may hint of a need to modify the distance and structure of
the shell. Precise analysis of such ``non-stable'' features requires
detailed model fitting beyond the scope of this paper.

In the recent past, both the scenarios leading to shell-like
structures have been discounted. PDD models have been dismissed
because 3D deflagration models showed that the WD becomes unbound and
thus pulsations would not occur
\citep{roepke2003,hillebrandt2000}. However, it has recently been
shown that this solution depends mainly on the ignition conditions,
namely the number and locations of ignition points leading to single
or multiple bubble solutions, and mixture of bubble solutions leading
to Raleigh-Taylor instabilities. As a result, solutions with fewer
bubbles are likely to result in a reduced amount of burning, thus only
slightly unbinding the WD and increasing the possibility of PDDs
\citep{livne2005,plewa2004}. Similarly, the merging scenario has been
dismissed because the WD may undergo burning during the merger and
result in an accretion induced collapse \citep{nomoto1991}, and also
on the basis of the long merging time scale. However both of these
results depend sensitively on the initial conditions, and new pathways
to the actual merging may effect the results \citep{lu2006}. In light
of our results, the predicted death of both of these scenarios may be
premature, and further studies are needed.

\acknowledgments We would like to thank the staff of the Hobby-Eberly
Telescope and McDonald Observatory for their support and the ROTSE
collaboration. We give specific thanks to J. Caldwell, S. Odewahn,
V. Riley, B. Roman, S. Rostopchin, M. Shetrone, E. Terrazas, and
M. Villarreal for their skilled observations with the HET, and to
F. Castro, P. Mondol, and M. Sellers for their efforts in screening
potential SN candidates.  This work made use of the SUSPECT on-line
database of SNe spectra
(\url{http://bruford.nhn.ou.edu/$\sim$suspect/index1.html}).  This
research is supported, in part, by NASA grant NAG 5-7937 (PH) and NSF
grants AST0307312 (PH) and AST0406740 (RQ \& JCW).

\clearpage

\begin{deluxetable}{cccccc}
\tablewidth{0pt}
\tablecaption{Observing Log for HET Spectra of SN~2005hj}
\tabletypesize{\scriptsize}
\tablehead{
  \colhead{Date (UT)} &
  \colhead{JD-2400000.5} &
  \colhead{Phase (day)} &
  \colhead{Exp. (s)} &
  \colhead{Filter} &
  \colhead{Slit ($''$)}
}
\startdata
Oct 26.30 &   53669.30 &      -6 & 2x600 &   GG385    &   2.0  \\
Oct 27.20 &   53670.20 &      -5 & 4x600 &   OG590    &   1.5  \\
Oct 27.28 &   53670.28 &      -5 & 2x600 &   GG385    &   1.5  \\
Nov  1.27 &   53675.27 &       0 & 2x600 &   GG385    &   1.5  \\
Nov  1.29 &   53675.29 &       0 & 2x550 &   OG515    &   1.5  \\
Nov  3.27 &   53677.27 &       2 & 2x480 &   GG385    &   1.5  \\
Nov  4.26 &   53678.26 &       3 & 2x600 &   OG515    &   1.5  \\
Nov  6.17 &   53680.17 &       5 & 2x600 &   GG385    &   1.5  \\
Nov  6.18 &   53680.18 &       5 & 2x600 &   OG515    &   1.5  \\
Nov 11.25 &   53685.25 &      10 & 3x600 &   OG515    &   1.5  \\
Nov 19.22 &   53693.22 &      18 & 4x600 &   OG515    &   1.5  \\
Nov 26.20 &   53700.20 &      25 & 2x600 &   OG515    &   1.5  \\
Dec  5.18 &   53709.18 &      34 & 3x600 &   OG515    &   2.0  \\
\enddata

\tablecomments{Phases are relative to the derived date of maximum
light, 2005 Nov. 1.6, and rounded to the nearest day. Effective
wavelength ranges for the GG385, OG515, and OG590 setups are
4100-8900~\AA, 5100-9800~\AA, and 5900-10,800~\AA, respectively. All
spectra from a given night were combined for the figures and
analysis.}

\label{specobs}
\end{deluxetable}

\clearpage

\begin{deluxetable}{rrrrr}
\tablewidth{0pt}
\tablecaption{Derived Quantities for the SN~2005hj \ion{Si}{2} $\lambda$6355 Line}
\tabletypesize{\scriptsize}
\tablehead{
  \colhead{Phase} &
  \colhead{Velocity} &
  \colhead{$\sigma_{\rm{vel}}$} &
  \colhead{Depth$^{a}$} &
  \colhead{FWHM} \\

  \colhead{(days)} &
  \colhead{(km s$^{-1}$)} &
  \colhead{(km s$^{-1}$)} &
  \colhead{} &
  \colhead{(\AA)}
}
\startdata
    $-6$ &    10820 &      140 &     0.38 &      140  \\
    $-5$ &    10800 &      110 &     0.32 &      120  \\
       0 &    10640 &       90 &     0.52 &      110  \\
       2 &    10440 &      100 &     0.60 &      110  \\
       3 &    10640 &       90 &     0.57 &      110  \\
       5 &    10680 &       80 &     0.57 &      110  \\
      10 &    10530 &      100 &     0.60 &      100  \\
      18 &    10550 &      120 &     0.48 &       90  \\
      25 &     9850 &       90 &     0.25 &       60  \\
      34 &  \nodata &  \nodata & \nodata  & \nodata   \\
\enddata

\tablecomments{Quantities measured from the FT smoothed data. The
\ion{Si}{2} $\lambda$6355 line is not clearly detected in the $+34$
day spectrum and thus no measurements are reported.}
\tablenotetext{a} {Similar to \citet{leonard2002}, depth is defined as
$(f_c - f_{\rm{min}})/f_c$ where $f_{\rm{min}}$ is the flux at the
minimum of the smoothed line and $f_c$ is the estimated continuum
level at the corresponding wavelength.}

\label{linedata}
\end{deluxetable}

\clearpage

\clearpage

\clearpage

\clearpage

\clearpage

\clearpage

\clearpage

\clearpage

\clearpage

\clearpage

\end{document}